\documentclass[11pt]{article} \hoffset=-15mm \voffset=-10mm
\usepackage{graphicx,amsmath,amssymb,epsf} \usepackage{latexsym,bm,
  slashed} \usepackage{xcolor} \definecolor{dark}{rgb}{0.10,0.2,0.3}
\definecolor{magenta}{rgb}{0.7,0.1,0.3}
\definecolor{purpure}{rgb}{0.5,0.15,0.3}
\usepackage[font=small,format=plain,labelfont=bf,up,textfont=it,up]{caption}
\usepackage{hyperref, cite} \hypersetup{colorlinks, citecolor=blue,
  filecolor=blue, linkcolor=magenta,
  urlcolor=purpure,hyperfootnotes=true,pdftex}

 \title{\bf  \Large 
QCD evolution based evidence for the onset of gluon saturation in exclusive photo-production of vector mesons
   } 
\author{ A.~Arroyo~Garcia${}^{a}$, M.~Hentschinski${}^{a}$,
  K.~Kutak${}^{b}$\\ \\
${}^{a}$ Departamento de Actuaria, F\'isica y Matem\'aticas,
Universidad de las Americas Puebla, \\ Santa Catarina Martir, 72820 Puebla, Mexico\\
${}^{b}$ The H. Niewodnicza\'nski Institute of Nuclear Physics, \\  Polish Academy of Sciences,ul. Radzikowskiego 152, 31-342, Cracow, Poland
}

\begin{document}

\maketitle
\begin{abstract}
  We investigate photo-production of vector mesons $J/\Psi$ and
  $\Upsilon$ measured  both at  HERA and LHC, using 2 particular fits
  of inclusive unintegrated gluon distributions, based on non-linear
  Balitsky-Kovchegov evolution (Kutak-Sapeta gluon; KS) and
  next-to-leading order Balitsky-Fadin-Kuraev-Lipatov evolution
  (Hentschinski-Sabio Vera-Salas gluon; HSS). We find that linear
  next-to-leading order evolution can only describe production at
  highest energies, if perturbative corrections are increased to
  unnaturally large values; rendering this corrections to a
  perturbative size, the growth with energy is too strong and the
  description fails. At the same time, the KS gluon, which we explore
  both with and without non-linear corrections, requires the latter to
  achieve an accurate description of the energy dependence of data. We
  interpret this observation as a clear signal for the presence of
  high gluon densities in  the proton, characteristic for the onset
  of gluon saturation.
\end{abstract}

\section{Introduction}
\label{sec:intro}

Energies available at the LHC allow for a detailed study of dynamical
effects of Quantum Chromodynamics (QCD). A prominent example is the
Heavy Ion program which focuses on the study of the Quark Gluon Plasma
(QGP). While many of the QGP properties are understood, the initial
state leading to its formation poses still open questions
\cite{Gelis:2015gza,Gelis:2010nm}. A closely related question is the
formation of an over occupied system of gluons, which eventually leads
to saturation of gluon densities \cite{Gribov:1984tu}; finding
convincing phenomenological evidence for gluon saturation is still one
of the open problems of QCD. On microscopic level, gluon saturation is expected to arise as a consequence of  recombination
of so-called wee gluons. The net effect of such recombination is to
slow down the growth of gluon number density with energy, commonly referred to as
gluon saturation.  The evolution from the low to large gluon densities
is described by a set of nonlinear evolution equations, known as
Balitsky-Jalilian-Marian-Iancu-McLerran-Weigert-Leonidov-Kovner
evolution
\cite{Balitsky:1995ub,JalilianMarian:1997jx,JalilianMarian:1997gr};
its frequent used mean field version is given by the Balitsky
Kovchegov (BK) \cite{Balitsky:1995ub,Kovchegov:1999yj} evolution
equation. First hints for gluon saturation could be already found  in
data collected from HERA experiments \cite{GolecBiernat:1998js},
(for a review see: \cite{Kovchegov:2012mbw} and references
therein). More recent studies focus on resolving hadronic final states
 at RHIC and LHC to explore the transverse momentum distribution in the
 potentially saturated proton and nucleus, {\it e.g.} hadron
production \cite{Albacete:2010pg,Albacete:2014fwa,Lappi:2012nh,Stasto:2018rci,Albacete:2018ruq} or
azimuthal de-correlations of dijets \cite{vanHameren:2019ysa}. Such
studies allow to study signals of saturation linked indirectly to the
energy dependence of gluon distributions. \\

In contrast to these studies, exclusive production of vector mesons
provides a direct opportunity to explore the energy dependence of the
cross section \cite{Armesto:2014sma,Goncalves:2014wna,Goncalves:2014swa}.  Since for
this observable large amount of data on production of
$J/\Psi, \Upsilon(1S)$ has been collected over wide range of energies,
it allows to potentially observe the predicted slow down of the growth
with energy of cross-sections, which is one of the core predictions
associated with the presence of high and saturated gluon densities.
Studies in the literature for this process, which take into account
effects due to gluon saturation, exist  both on the level of
dipole models \cite{Armesto:2014sma,Goncalves:2014wna,Goncalves:2014swa,Kowalski:2006hc,Cox:2009ag}
and complete solutions to non-linear BK equation
\cite{Ducloue:2016pqr,Cepila:2018faq}. The description of the cross
sections is in general satisfactory. Here we would like to
re-investigate the problem from a different angle: While the ability of
non-linear evolution equations to describe collider data is a
necessary requirement to establish the presence of gluon saturation,
it is not sufficient: albeit a successful description of data is
provided, one might still remain in the so-called dilute regime, where
gluon densities are perturbative. That this could be actually the
case, is at least suggested by the successful description of the same
data set by frameworks which rely on collinear factorization
\cite{Jones:2013eda,Jones:2013pga,Jones:2016icr,Szczurek:2017uvc} and
-- maybe even more striking -- linear NLO BFKL evolution
\cite{Bautista:2016xnp}. The latter relies on a fit of an initial (low
energy) transverse momentum distribution to combined HERA data by
Hentschinski-Salas-Sabio Vera (HSS)\cite{Hentschinski:2012kr}, and has
been recently explored in a number of phenomenological studies, see
{\it e.g.}  \cite{Celiberto:2018muu}.\\

In the following we will argue that even though a description of
exclusive photo-production data by linear QCD evolution can be
achieved for the entire range of available center-of-mass energies,
such a description requires unnaturally large higher order
corrections, which in some region of phase space are larger than the
dominant leading order term. To assess the importance of non-linear
terms in low $x$ evolution equations, we use a particular solution to
BK-evolution, with initial conditions fitted to combined HERA data by
Kutak-Sapeta (KS) \cite{Kutak:2012rf}. While the HSS-gluon relies on
NLO-BKFL evolution and KS-evolution on LO-BK evolution, both evolution
schemes supplement the original low $x$ evolution with a
supplementary resummation of collinear logarithms \cite{Kwiecinski:1997ee}. Our strategy is
then as follows: While exclusive photo-production of the $\Upsilon$
serves as the necessary cross-check in the perturbative, dilute region
(provided by the relatively large bottom quark mass), saturation
effects will be searched for in photo-production of the $J/\Psi$
mesons (where the hard scale is provided by the charm quark mass). In
particular we will demonstrate that the seemingly flawless description
of $J/\Psi$-data by NLO BFKL evolution is only possible due to the
presence of a very large perturbative correction, which slows down the
growth with energy. Choosing on the other hand a hard scale which
renders this perturbative corrections small, the data are no longer
described. The KS gluon (which is subject to non-linear QCD evolution)
is on the other hand able to describe the energy dependence of
$J/\Psi$-data. Turning off the non-linear effects in the evolution of
the KS-gluon, we find that a description of the energy dependence is
no longer possible. We are convinced that this observation provides a
very strong phenomenological evidence for the presence of saturation
effects in the high $W$ region of $J/\Psi$ photo-production observed
at the LHC.

The outline of this letter is as follows: In Sec.~\ref{sec:Xsection}
we provide details of our theoretical description,
Sec.~\ref{sec:pert_corrections} is dedicated to a discussion of the
large perturbative corrections of the NLO BFKL gluon in the large $W$
region while in Sec.~\ref{sec:concl} we present our conclusions.

\section{Energy dependence of the photo-production cross-section}
\label{sec:Xsection}
 We study the process \footnote {Besides HERA data we also use the LHC p-p and Pb-p data where highly boosted p and Pb respectively become a source of photons leading to Ultra Peripheral Collisions}
$ \gamma(q) + p(p)  \to V(q') + p(p')  $ 
where $V = J/\Psi, \Upsilon(1S)$ and $\gamma$ denotes a quasi-real
photon with virtuality $Q \to 0$; $W^2 = (q + p)^2$ is the squared
center-of-mass energy of the $\gamma(q) + p(p)$
collision. The $x$ value probed in such a collision is obtained as $x \simeq M_V^2/W^2$ with $M_V$ the mass of the vector meson.   With the momentum transfer
$t = (q-q')^2$, the differential cross-section for
the exclusive production of a vector meson can be written in the
following form
\begin{align}
  \label{eq:16}
  \frac{d \sigma}{d t} \left(\gamma p \to V p \right)
& =
\frac{1}{16 \pi} \left|\mathcal{A}_{T, L}^{\gamma p \to V p}(W^2, t) \right|^2 \, ,
\end{align}
where $\mathcal{A}(W^2, t)$ denotes the scattering amplitude for the
reaction $\gamma p \to V p$ for color singlet exchange in the
$t$-channel, with an overall factor $W^2$ already extracted. For a more detailed discussion of the kinematics we refer to \cite{Bautista:2016xnp}. 

\subsection{The theoretical setup of our study}
\label{sec:theor-setup-our}

In the following we  determine the total photo-production
cross-section, based on an inclusive gluon distribution. This is possible
following a two step procedure, frequently employed in the
literature: First one determines the differential cross-section at zero
momentum transfer $t=0$ (which can be expressed in terms of the
inclusive gluon distribution); in a second step  the
$t$-dependence is modeled  which then allows us to relate  the
differential cross-section at $t=0$ to the integrated
cross-section. Here we follow the prescription given in
\cite{Jones:2013eda,Jones:2013pga}, where  an exponential drop-off
with $|t|$, $\sigma \sim \exp\left[-|t| B_D(W)\right]$ is used  with an energy dependent
$t$ slope parameter $B_D$, as motivated by Regge theory,
\begin{align}
  \label{eq:18}
  B_D(W) & =\left[  b_0 + 4 \alpha' \ln \frac{W}{W_0} \right] \text{GeV}^{-2}.
\end{align}
Following \cite{Jones:2013eda,Jones:2013pga}, we use for the numerical values  $\alpha' = 0.06$ GeV$^{-2}$,
$W_0 = 90$ GeV and $b_0^{J/\Psi} = 4.9$ GeV$^{-2}$ in the case of the
$J/\Psi$, while $b_0^{\Upsilon} = 4.63$ GeV$^{-2}$ for $\Upsilon$
production.  The
total cross-section for vector meson production is therefore obtained
as
\begin{align}
  \label{eq:16total}
 \sigma^{\gamma p \to V p}(W^2) & = \frac{1}{B_D(W)} \frac{d \sigma}{d t} \left(\gamma p \to V p \right)\bigg|_{t=0}
.
\end{align}
The  uncertainty introduced by the modeling of
the $t$-dependence mainly affects the
overall normalization of the cross-section with a mild logarithmic
dependence on the energy. To determine the scattering amplitude, we
first note that the dominant contribution is provided by its imaginary
part. Corrections due to the real part of the scattering amplitude
can be estimated using dispersion relations, in particular 
\begin{align}
  \label{eq:32}
  \frac{\Re\text{e} \mathcal{A}(W^2, t)}{\Im\text{m} \mathcal{A}(W^2, t)}
&=
\tan \frac{\lambda \pi }{2},
& \text{with}&
& \lambda & = \frac{d \ln \Im\text{m}  \mathcal{A}(W^2, t) }{ d \ln W^2} \, .
\end{align}
As noted in \cite{Bautista:2016xnp,Baranov:2007zza}, the dependence of the slope
parameter $\lambda$ on energy $W$ provides a sizable correction to
the  $W$ dependence of the complete
cross-section. We therefore do not assume $\lambda =$
const., but instead determine the slope $\lambda$ directly from the $W$-dependent
imaginary part of the scattering amplitude. The latter is obtained from \cite{Cox:2009ag,
  Kowalski:2006hc}
\begin{align}
\label{am-i}  
\Im \text{m}\mathcal{A}^{\gamma p\rightarrow
    Vp}_{T} (W, t=0) &= 2 \, \int\!d^2{\bm r}\int\! d^2{\bm b}
  \int_0^1\! \frac{d {z}}{4 \pi }\;(\Psi_{V}^{*}\Psi)_{T}
  \;\mathcal{N}\left(x,r,b\right)\,,
\end{align}
where $\mathcal{N}\left(x,r,b\right)$ is the dipole amplitude and
$T$ denotes transverse  polarization of the quasi-real
photon.  Here
\begin{align}
  \label{eq:21}
  \left(\Psi_{V}^{*}\Psi\right)_{T}(r,z) & =  \frac{\hat{e}_f eN_c}{\pi z (1-z)}
 \bigg\{
 m_f^2 K_0(\epsilon r) \phi_T(r,z) - \left[z^2 + (1-z)^2 \right] \epsilon K_1(\epsilon r) \partial_r \phi_T(r,z)
\bigg\}
 \, ,
\end{align}
  with $\epsilon^2 = m_f^2$ for
real photons. Furthermore $r = \sqrt{{\bm r}^2}$, while $f = c, b$
denotes the flavor of the heavy quark and  $\hat{e}_f=2/3$, $-1/3$.  For the
scalar parts of the wave functions $\phi_{T,L}(r,z)$, we employ the boosted Gaussian
wave-functions with the Brodsky-Huang-Lepage prescription
\cite{Brodsky:1980vj}.  For the ground state vector meson ($1s$) the
scalar function $\phi_{T}(r,z)$, has the following general form
\cite{Cox:2009ag, Nemchik:1994fp},
\begin{align}
\label{eq:1s_groundstate}
\phi_{T,L}^{1s}(r,z) &= \mathcal{N}_{T,L} z(1-z)
  \exp\left(-\frac{m_f^2 \mathcal{R}_{1s}^2}{8z(1-z)} - \frac{2z(1-z)r^2}{\mathcal{R}_{1s}^2} + \frac{m_f^2\mathcal{R}_{1s}^2}{2}\right)  \, .
\end{align}
The free parameters $N_T$ and $\mathcal{R}_{1s}$ of this model have been
determined in various studies from the wave function normalization  and the decay width of the vector mesons. In the
following we use the values found in 
\cite{Armesto:2014sma} ( $J/\Psi$) and \cite{Goncalves:2014swa}
( $\Upsilon$). The parameters are summarized in Tab.~\ref{vm_fit}.
\begin{table}
\centering
\begin{tabular}{c|c|c|c|c}
\hline\hline &&&& \vspace{-.2cm}\\
Meson & $m_f/\text{GeV}$  & $\mathcal{N}_T$ &  $\mathcal{R}^2$/$\text{GeV}^{-2}$
  & $M_V$/GeV     \\
&&&& \vspace{-.2cm}\\  \hline
$J/\psi$ & $m_c=1.4$&   $0.596$ & $2.45$ & $3.097$    \\ \hline
$\Upsilon$ & $m_b = 4.2 $ & $0.481$ & $0.57$ & $9.460$  \\ \hline

\end{tabular}
\caption{Parameters of the boosted Gaussian vector meson wave functions for $J/\psi$ and $\Upsilon$  \cite{Armesto:2014sma, Goncalves:2014swa}.}
\label{vm_fit}
\end{table}
In the forward limit $t=0$, one further has,
\begin{align}
  \label{eq:intN}
  2 \int d^2 {\bm b} \,   \mathcal{N}\left(x,r,b\right) & = \sigma_{q\bar{q}}(x, r) \,,
\end{align}
where $\sigma_{q\bar{q}}$ denotes the  inclusive dipole
cross-section which is related to the   unintegrated gluon density $ {\cal F}$
through  \cite{Braun:2000wr} 
\begin{align}
  \label{eq:Nfromugd}
 \sigma_{q\bar{q}}(x, r) & = \frac{4 \pi}{N_c} \int \frac{d^2 {\bm
                           k}}{{\bm k}^2}
\left(1-e^{i {\bm k}\cdot {\bm r}}\right) \alpha_s  {\cal F}(x, {\bm k}^2) \, .
\end{align}
 In \cite{Bautista:2016xnp} this expression has  been used to
 calculate the BFKL impact factor  in transverse Mellin space from the
 light-front wave function overlap Eq.~\eqref{eq:21}. In the following
 study we chose a slightly different route and calculate the dipole
 cross-section directly from the regarding KS and HSS unintegrated gluon
 densities. These gluon densities have been obtained in the following way:
\begin{itemize}
\item The KS gluon is obtained as  a solution of the momentum space version of BK equation with modifications
according to the Kwieciński-Martin-Satsto (KMS) prescription
\cite{Kwiecinski:1997ee} which includes a kinematical constraint to
impose  energy momentum conservation, complete DGLAP splitting function
and contribution of quarks as well as the  1-loop QCD running
coupling. As a consequence  this evolution equation reduces in the
collinear limit  to leading order  DGLAP evolution. The intial
conditions of the KS gluon distribution  have been determined using a
fit  \cite{Kutak:2012rf}  combined HERA  data for the proton structure
function $F_2$  \cite{Aaron:2009aa} in the region $x < 10^{-2}$. 
\item Below we will further consider the linear version of the KS
  gluon which is obtained as a solution of momentum space version of
  the leading order BFKL equation with
modifications according to the KMS prescription described above. The
initial conditions of the linear equation have been obtained from a
fit  to HERA data, similar to the non-linear case, but with photon
virtualities restricted to the region $Q^2 > 4.5$~GeV$^2$. 
\item The HSS gluon is obtained as a solution to the NLO BFKL equation
  in transverse Mellin space applying both a resummation of  collinear
  logarithms within the NLO BFKL kernel and a resummation of large
  running coupling corrections (optimal renormalization scale
  setting). The initial conditions have been fitted
  \cite{Hentschinski:2012kr} to combined HERA data for  proton
  structure function $F_2$ in the region $x < 10^{-2}$ and photon
  virtualiyt $Q^2 \geq 1.2$~GeV$^2$. 
\end{itemize}
 For a detailed discussion of the framework underlying both
 gluon distributions we refer to the literature:
 \cite{Kutak:2012rf,Kwiecinski:1997ee} (KS) and
 \cite{Hentschinski:2012kr,Chachamis:2015ona} (HSS). 
\begin{figure}[p!]
  \centering
  \includegraphics[width=.95\textwidth]{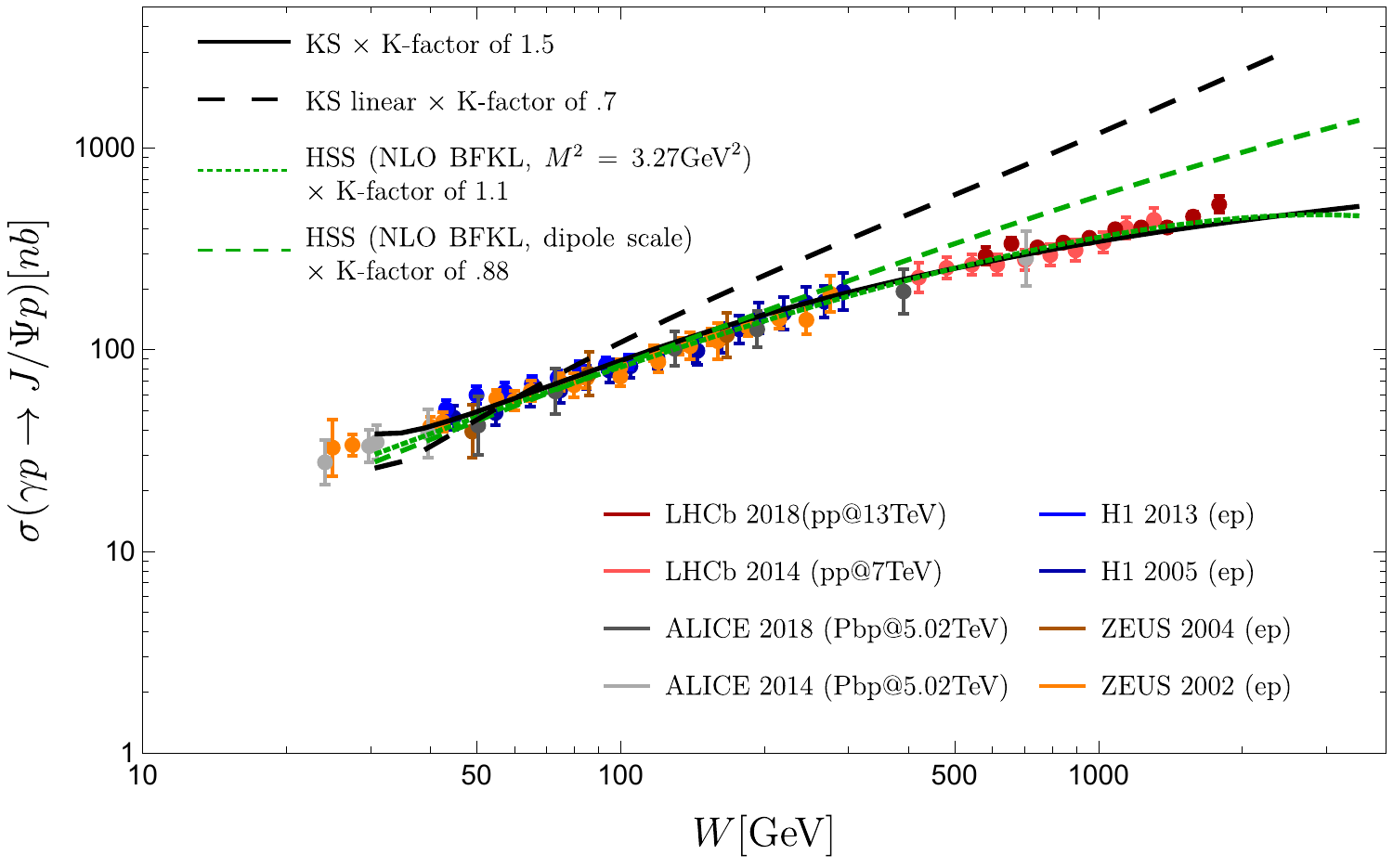}\\
\vspace{1cm}
  \includegraphics[width=.95\textwidth]{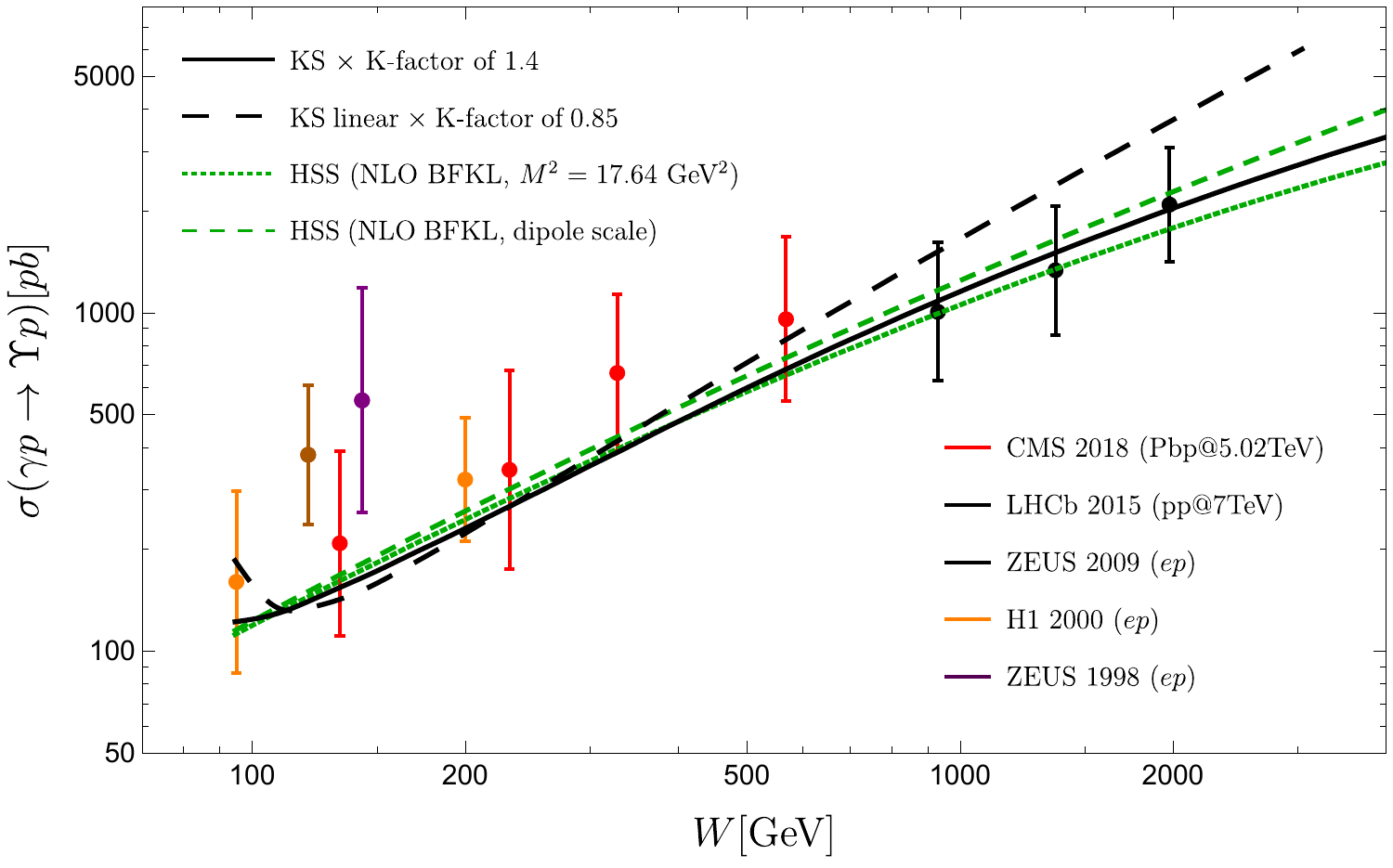}
  \caption{\it Energy dependence of the $J/\Psi$ and $\Upsilon$
    photo-production cross-section as provided by the KS and HSS gluon
    distribution. The HSS distribution with dipole size scale
    corresponds to a specific scale setting for the HSS gluon
    discussed in Sec.~\ref{sec:pert_corrections}.  For the $J/\Psi$ we
    further display photo-production data measured at HERA by ZEUS
    \cite{Chekanov:2002xi,Chekanov:2004mw} and H1
    \cite{Alexa:2013xxa,Aktas:2005xu} as well as LHC data obtained
    from ALICE \cite{TheALICE:2014dwa} and LHCb ($W^+$ solutions)
    \cite{Aaij:2013jxj,Aaij:2018arx}.  For the $\Upsilon$ cross-section we show
    HERA data measured by H1 \cite{Adloff:2000vm} and ZEUS
    \cite{Breitweg:1998ki,Chekanov:2009zz} and LHC data by LHCb
    \cite{Aaij:2015kea} and CMS \cite{CMS:2016nct,Sirunyan:2018sav}.}
  \label{fig:results}
\end{figure}

\subsection{Numerical results using standard implementations}
\label{sec:numer-results-using}

The main uncertainty left is the scale  at which the strong
 coupling constant $\alpha_s$ is to be evaluated in
 Eq.~\eqref{eq:Nfromugd}. In the case of the $J/\Psi$, which is
 characterized by a relatively small hard scale $m_c \simeq  1.4$~GeV and therefore large
 value of the strong coupling constant $\alpha_s \simeq 0.31$, this 
 leads to a sizable ambiguity in the normalization of the total
 cross-section, since the latter depends through Eq.~\eqref{eq:16}
 quadratically on $\alpha_s$. Using similar  conventions
 as used in original fits of the KS
 and HSS gluon distributions, we fix this scale
 to a typical hard scale of the process. For the KS gluon we chose
 both for the photo-production of  $\Upsilon$ and $J/\Psi$ vector
 mesons, the mass of the respective heavy quark as the hard
 scale; for the HS-gluon  it was found in
 \cite{Bautista:2016xnp}  that a scale related to
 the size of the $J/\Psi$ wave function is more suitable  in the case of  $J/\Psi$-production,  $M_{J/\Psi}^{\text{HS}} = 8/\mathcal{R}_{J/\Psi}^2= 3.27$~GeV while we use the bottom quark mass for $\Upsilon$-production.  The results of our study for the fixed scale case can be found in
 Fig.~\ref{fig:results}, where continuous, black lines correspond to
 the KS-gluon and dotted, green lines to the HS-gluon at fixed scales;
 the dashed green lines, corresponding to a special scale setting of
 the HSS gluon, and the linear KS-gluon will be discussed in the
 forthcoming section. We observe that both the KS-gluon distribution
 and the HSS-gluon distribution provide an excellent description of
 the energy dependence of the data. While the KS-gluon requires in the
 current study a K-factor of $1.4-1.5$, we note that the size of such
 a correction strongly depends on the precise scale choice of the
 strong coupling constant and the precise parametrization used for the
 $t$-slope parameter $B_D$, Eq.~\eqref{eq:18}. Indeed, using the
 parametrization of the $B_D$ parameter suggested in
 \cite{Goncalves:2014swa}, would bring the K-factor of the KS-gluon
 close to one in the case of $\Upsilon$-photo-production. We further
 note that our study does not include a frequently employed
 phenomenological corrective factor which can be determined through
 relating the inclusive collinear gluon distribution to generalized
 parton distribution through a Shuvaev transform. While the precise
 numerical value of this factor depends on the energy dependent slope
 parameter $\lambda$,  it generally provides a
 correction which is of the order of the K-factor found for the
 KS-gluon.  Our general conclusion is that the theoretical accuracy of
 the observable is currently not sufficient to fix unambiguously the
 normalization. At the same time, the energy dependence is described
 in an excellent way, both by KS-evolution (non-linear BK evolution
 combined with DGLAP corrections and kinematical constraint and
 collinear resummation) and the HSS-gluon (NLO BFKL evolution with
 collinear resummation).

\section{The need for non-linear low $x$ evolution}
\label{sec:pert_corrections}
At first sight it appears that both non-linear $W$ evolution (KS) and
linear $W$ evolution (HSS) describe the $W$-dependence of data; one
might be therefore lead to conclude that non-linear effects,
associated with the presence of high and saturated gluon densities,
are absent and the observables merely probes the linear, dilute
regime. In the following we argue that such a conclusion is
pre-mature. Indeed there are strong hints which suggest that we
are at least in the transition region towards high and saturated gluon
densities. \\

To fully access this question, we first recapitulate which possible
impact large gluon densities could have on the observable. First of
all, the presence of high density effects cannot be seen directly at
the level of the observable. The scattering amplitude Eq.~\eqref{am-i}
depends only on the dipole amplitude, which itself can be expressed as
the correlator of two Wilson lines which resum the gluonic field of
the proton, see {\it e.g.} \cite{Weigert:2005us}. Even though the
dipole amplitude resums the interaction of the $q\bar{q}$-dipole with
an infinite number of gluons, the gluons couple to the
$q\bar{q}$-dipole like a single gluon; the ``reggeize'' in the
language of \cite{Bartels:1994jj} and therefore appear like a single
gluon. At the level of our phenomenological study, this property
reveals itself through Eq.~\eqref{eq:Nfromugd}, which relates the
dipole cross-section to the unintegrated gluon density. To make
multiple re-scattering of partons on the target field visible, it
would be necessary to resolve the hadronic final state of the
dissociated photon, see {\it e.g.}
\cite{Ayala:2017rmh,Kotko:2017oxg}. This not the case for
photo-production of vector mesons.  The only place where one could
expect a signal for the presence of saturation effects is therefore
the $x$-dependence of the underlying gluon distributions. As an
immediate consequence, any framework which is based on a direct fit of
the $x$-dependence at the $J/\Psi$ scale (such as collinear parton
distribution functions) does not exclude presence of saturation
effects; it merely demonstrates the ability to fit the resulting
$x$-dependence of the underlying gluon distribution. While this
initial $x$-distribution can be evolved through DGLAP evolution to
events with higher hard scales, such events are generally
characterized by larger values of $x$
($x_{\Upsilon} > 2.28 \cdot 10^{-5}$ vs.
$x_{J/\Psi} > 2.99 \cdot 10^{-6}$ in the current case).  Taking
further into account that DGLAP evolution is known to shift large $x$
input to lower $x$, it is therefore safe to say that the mere ability
of DGLAP fits to accommodate low $x$ $J/\Psi$ photo-production data,
does not
exclude the potential presence of sizable non-linear effects for the data points at highest $W$-values. \\

Instead of DGLAP evolution, a suitable benchmark to establish presence/absence of gluon saturation is  provided by linear NLO BFKL 
evolution, such as the HSS gluon.  While the HSS gluon provides a very good
description of both $\Upsilon$ and $J/\Psi$ photo-production data, the
following observation can be made: Recalling the  particularly solution of NLO
BFKL evolution used for the HSS-fit, one finds at the  at level of the dipole
cross-section two terms
\begin{align}
  \label{eq:1}
  \sigma_{q\bar{q}}^{(\text{HSS})} (x, r) & = \alpha_s   \hat{\sigma}_{q\bar{q}}^{(\text{HSS})} (x, r) ,
&
\hat{\sigma}_{q\bar{q}}^{(\text{HSS})} (x, r) & =  \hat{\sigma}_{q\bar{q}}^{(\text{dom.})} (x, r)  +   \hat{\sigma}_{q\bar{q}}^{(\text{corr.})} (x, r)  ,
\end{align}
where 
\begin{align}
  \label{eq:2}
\hat{\sigma}_{q\bar{q}}^{(\text{dom})} (x, r, M^2)   &= \int\limits_{\frac{1}{2} - i \infty} ^{\frac{1}{2} + i \infty}\!\! \frac{d \gamma}{2 \pi i} \left(\frac{4}{r^2 Q_0^2} \right)^\gamma \frac{ \bar{\alpha}_s(M \cdot Q_0)}{\bar{\alpha}_s(M^2)}
 f(\gamma, Q_0, \delta, r)   \left(\frac{1}{x}\right)^{\chi\left(\gamma, M^2 \right)}\notag \\
\hat{\sigma}_{q\bar{q}}^{(\text{corr.})} (x, r, M^2)   &= \int\limits_{\frac{1}{2} - i \infty} ^{\frac{1}{2} + i \infty}\!\! \frac{d \gamma}{2 \pi i} \left(\frac{4}{r^2 Q_0^2} \right)^\gamma \frac{ \bar{\alpha}_s(M \cdot Q_0)}{\bar{\alpha}_s(M^2)}
 f(\gamma, Q_0, \delta, r)   \left(\frac{1}{x}\right)^{\chi\left(\gamma, M^2 \right)}\notag \\
&\hspace{-2cm} \times   \frac{\bar{\alpha}_s^2\beta_0  \chi_0 \left(\gamma\right) }{8 N_c} \log{\left(\frac{1}{x}\right)}
  \Bigg[- \psi \left(\delta-\gamma\right)
 +  \log \frac{M^2r^2}{4} - \frac{1}{1-\gamma} - \psi(2-\gamma) - \psi(\gamma) \Bigg]\;,
\end{align}
and
\begin{align}
  \label{eq:5}
  f(\gamma, Q_0, \delta, r)&= \frac{  r^2 \cdot \pi \Gamma(\gamma) \Gamma(\delta - \gamma) }{ N_c (1-\gamma) \Gamma(2-\gamma) \Gamma(\delta)},
\end{align}
is a function which collects both factors resulting from the proton impact factor and the transformation of the unintegrated gluon density to the dipole cross-section, see \cite{Hentschinski:2012kr,Bautista:2016xnp} for details. The parameters $Q_0 = 0.28$~GeV, $\mathcal{C}=2.29$ and $\delta = 6.5$ have been determined from the HERA data fit.  Furthermore  $\bar{\alpha}_s = \alpha_s N_c/\pi$ with $N_c$ the number of
colors, and  $\chi(\gamma,M^2)$ is the next-to-leading
logarithmic (NLL) BFKL kernel after collinear improvements; in
addition large terms proportional to the first coefficient of the QCD
beta function, $\beta_0 =  11 N_c/3 - 2 n_f /3$ have been resumed
through employing a Brodsky-Lepage-Mackenzie (BLM) optimal scale
setting scheme \cite{Brodsky:1982gc}.  The NLL kernel with collinear
improvements reads
 \begin{align}\label{eq:gluongf}
\chi\left(\gamma, M^2 \right) &=
{\bar\alpha}_s\chi_0\left(\gamma\right) +
{\bar\alpha}_s^2\tilde{\chi}_1\left(\gamma\right)-\frac{1}{2}{\bar\alpha}_s^2
\chi_0^{\prime}\left(\gamma\right)\chi_0\left(\gamma\right)
+ \chi_{\text{RG}}({\bar\alpha}_s,\gamma,\tilde{a},\tilde{b})  .
\end{align}
where $\chi_i$, $i=0,1$ denotes the LO and NLO BFKL eigenvalue and
$\chi_{\text{RG}}$ resums (anti-)collinear poles to all orders; for
details about the individual kernels see
\cite{Hentschinski:2012kr,Bautista:2016xnp}.  The scale $M$ is a
characteristic hard scale of the process.  The second contribution
$\hat{\sigma}^{\text{corr.}}$ is proportional to $\beta_0$ and acts in
$\gamma$-space as a differential operator on the impact factors of
external particles. These terms do not exponentiate and have been
therefore treated in \cite{Hentschinski:2012kr} as a perturbative
correction to the BFKL Green's function.  Even though
$\hat{\sigma}^{\text{corr.}}$ is suppressed by a factor of
$\alpha_s^2$, enhancement by $\ln(1/x)$ will eventually compensate for
the smallness of the strong coupling constant and invalidate the
perturbative expansion.
\begin{figure}[t]
  \centering
 \parbox{.49\textwidth}{
 \includegraphics[width=.45\textwidth]{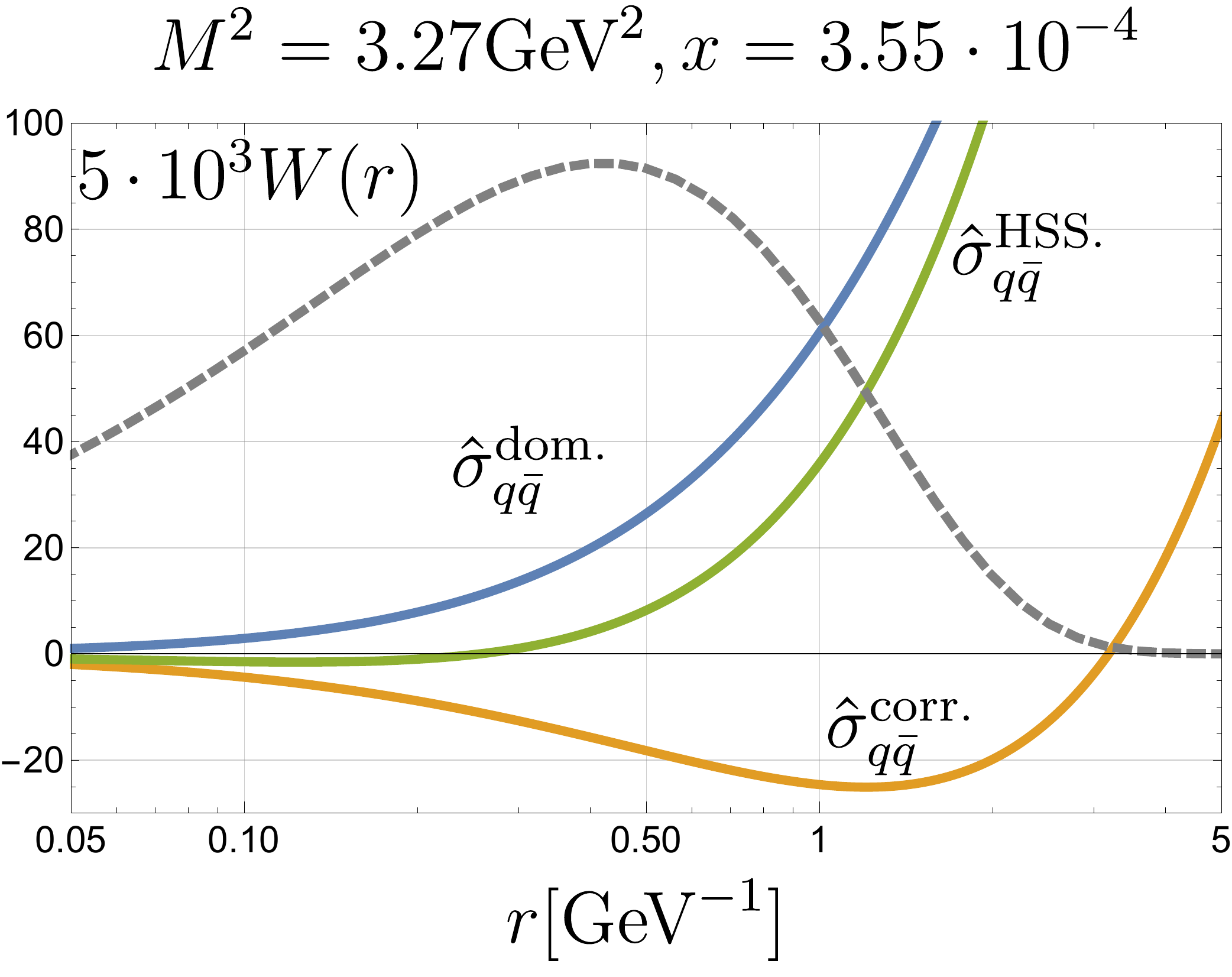}}
\parbox{.49\textwidth}{
  \includegraphics[width=.45\textwidth]{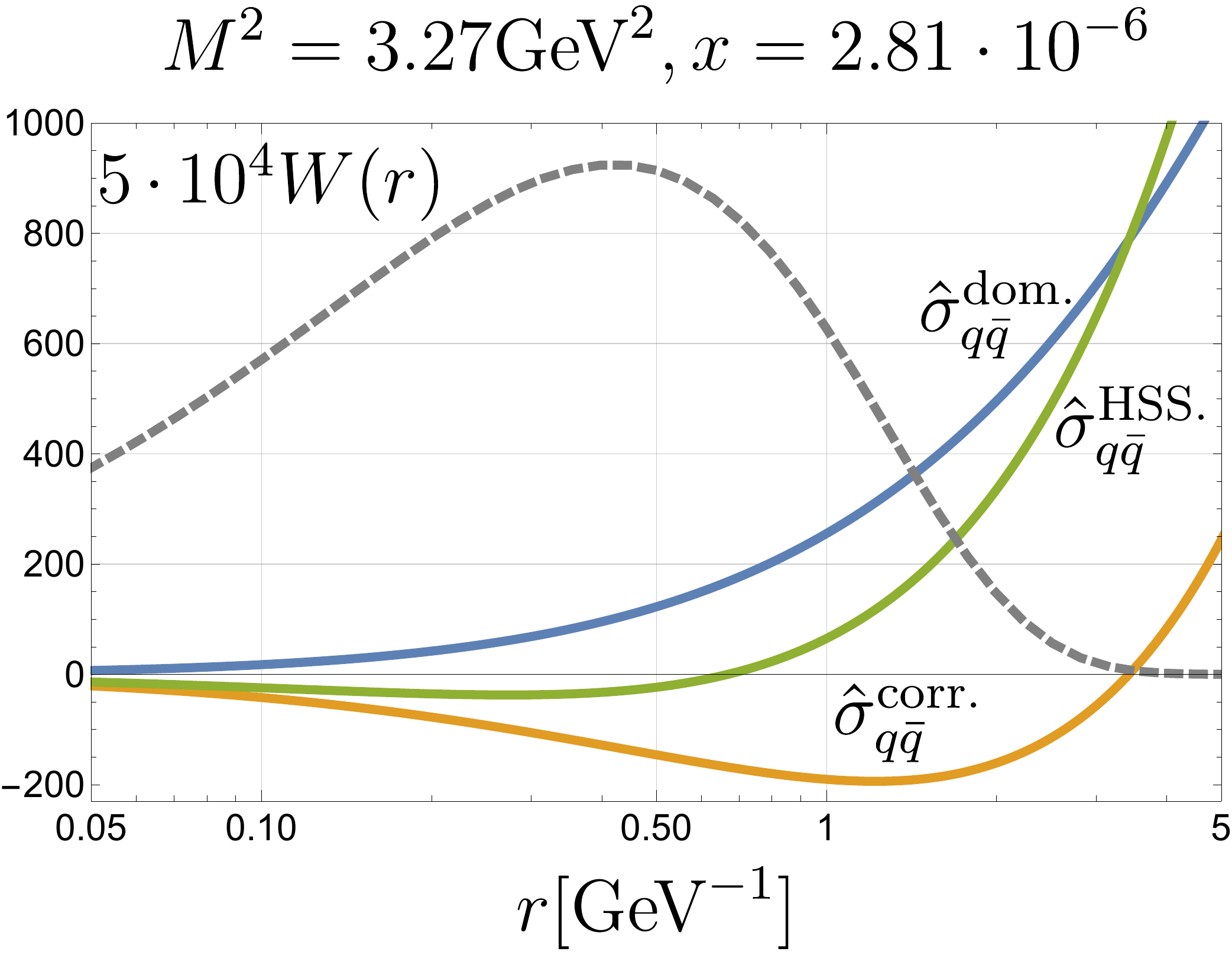}} \\
\parbox{.49\textwidth}{\center (a)}\parbox{.49\textwidth}{\center (b)}
 \parbox{.49\textwidth}{
   \includegraphics[width=.45\textwidth]{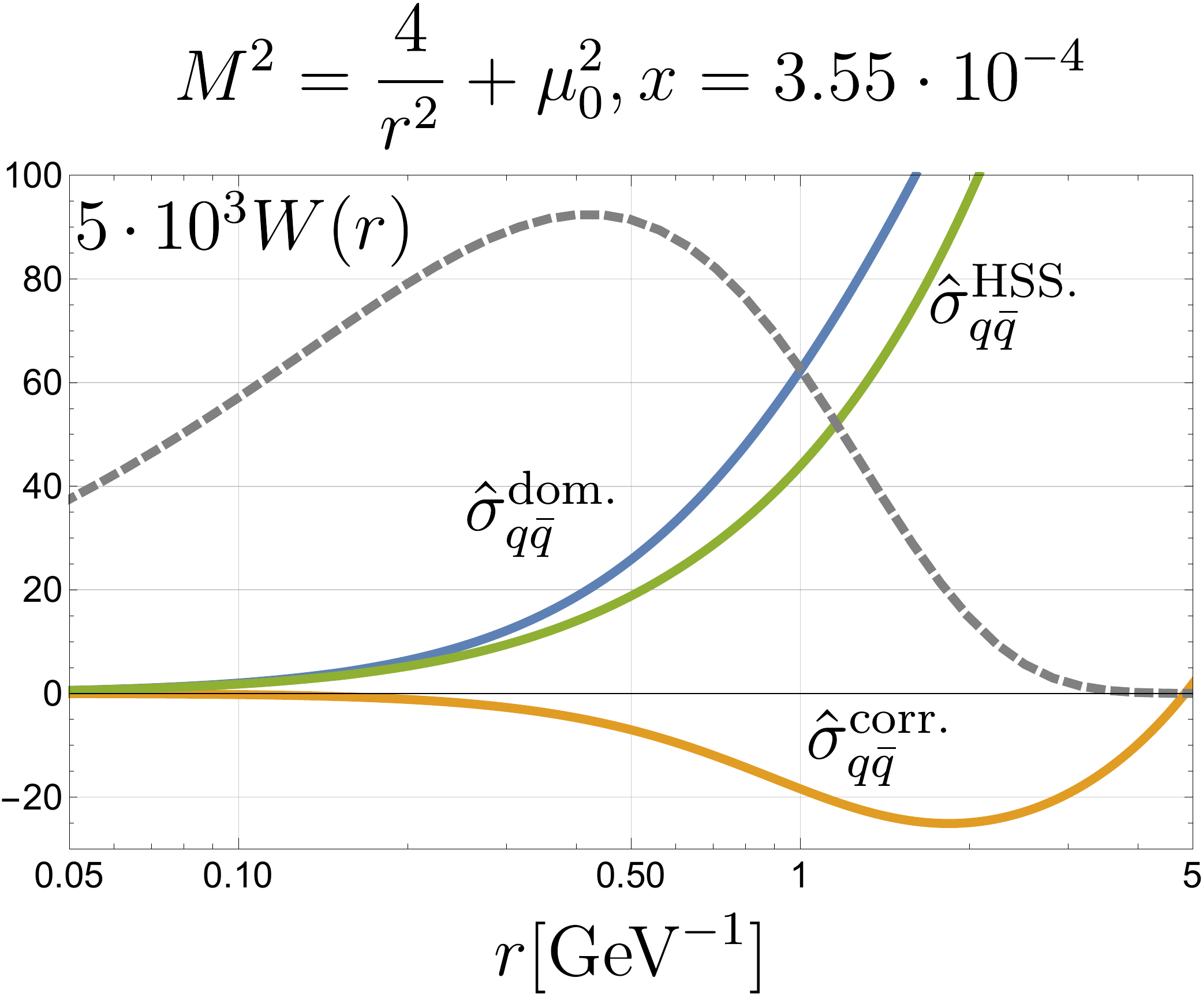} }
 \parbox{.49\textwidth}{
 \includegraphics[width=.45\textwidth]{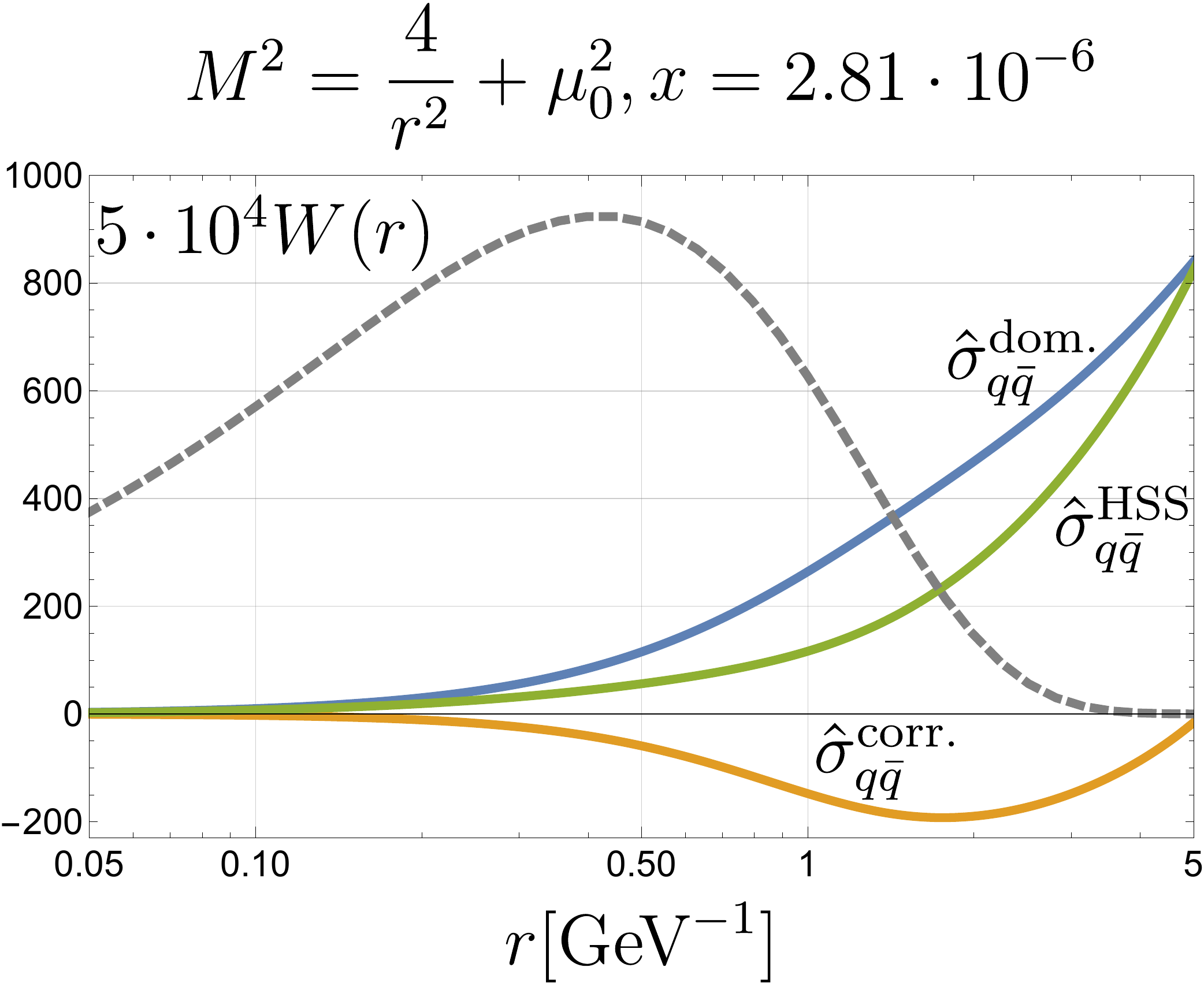}} \\
\parbox{.49\textwidth}{\center (c)}\parbox{.49\textwidth}{\center (d)}\\

  \caption{HSS dipole cross-section (with an overall factor of $\alpha_s$ extracted) at fixed ($J/\Psi$-scale, top row) and $r^2$-dependent scale (bottom row) in units of GeV$^{-2}$. We also show the function   $W(r)$ (times a suitable scaling factor) with which the dipole cross-sections are convoluted with to indicate the typical dipole sized probed in $J/\Psi$ photo-production. }
  \label{fig:HSS_dipole}
\end{figure}
The behavior of the HSS-dipole cross-section is studied in
Fig.~\ref{fig:HSS_dipole}. To  identify the relevant
region in dipole size $r$ for $J/\Psi$ photo-production we  further define
\begin{align}
  \label{eq:3}
  W(r) & =  2 \pi r \int_0^1\! \frac{d {z}}{4 \pi }\;(\Psi_{V}^{*}\Psi)_{T}(r,z),
\end{align}
as the $z$-integrated wave function overlap. Working with
a fixed hard scale $M^2 = 3.27$~GeV$^2$ we find in
Fig.~\ref{fig:HSS_dipole}, that the perturbative expansion is well
under control for a typical HERA $x$ value of $x = 3.55 \cdot 10^{-4}$
(Fig.~\ref{fig:HSS_dipole}.a). Turning however to the lowest $x$
values probed at the LHC of $x = 2.81 \cdot 10^{-6}$
(Fig.~\ref{fig:HSS_dipole}.b) we observe that the correction term is
generally large; for certain $r$ values,  which are further
enhanced through the $r$-dependence of $W(r)$,  they even super-seed  the
dominant term,  resulting into a negative dipole cross-section. While the dipole cross-section is not an observable, this
clearly indicates a breakdown of the perturbative expansion for dipole
sizes where the integrated wave function
overlap $W(r)$ has its maximum value. The problem of unnaturally large
higher order corrections can be fixed by choosing a hard scale related
to the transverse size of the dipole. Following the scale setting used
in fits  \cite{Rezaeian:2012ji} of the IP-sat model \cite{Bartels:2002cj}, we may
therefore choose $M^2 = \frac{4}{r^2} + \mu_0^2$ with
$ \mu_0^2 = 1.51~\text{GeV}^2$.  With this scale, we find  that
the perturbative expansion indeed stabilizes: both for the HERA
(Fig.~\ref{fig:HSS_dipole}.c) and LHC $x$-values
(Fig.~\ref{fig:HSS_dipole}.d) the perturbative term is well under
control. Turning with this choice for the hard scale however to data,
we find  that this scale setting (green dashed line in 
Fig.~\ref{fig:results})
describes very well the energy dependence of
$\Upsilon$-photo-production as well as $J/\Psi$ photo-production in the
HERA region $W < 300$~GeV, but  fails to describe  $J/\Psi$
production at the LHC ($W > 300$~GeV). The resulting  growth with
energy is too strong and  the data are no longer described
(Fig.~\ref{fig:results}, top). We therefore conclude that NLO BFKL
evolution can only describe data in the region $W > 300$~GeV if one
accepts very large perturbative corrections, which super-seed for certain
dipole sizes the dominant term and which slow down the growth of the cross-section. If the size of these perturbative
corrections is reduced using a suitable hard scale, the growth of the
HSS-gluon is too strong and cannot be accomodated by data.\\

The KS-gluon, which is subject to LO-BK evolution with collinear resummation provides a very good description of $J/\Psi$ data in the region  $W > 300$~GeV. To answer the question whether this description relies on the presence of non-linear terms in the evolution equation, we compare in addition to a linearized KS gluon (dashed black line in Fig.~\ref{fig:results}). We find that the growth in $W
$ is in that case even stronger than for the HSS gluon with dipole size scale. The ability of the KS gluon to describe data is therefore strongly connected to non-linear evolution effects.

\section{Conclusion and Discussion}
\label{sec:concl}

We conclude that there are strong hints for the presence of the
saturation effects in exclusive photo-production of $J/\Psi$ at small
$x$. While both linear and non-linear low $x$ QCD evolution can
describe the data, the former requires the presence of unnaturally
large perturbative corrections. Rendering these corrections small,
$\Upsilon$-data, characterized by a hard perturbative scale, are still
well described while the growth with energy is too strong for $J/\Psi$
data in the LHC region  with $W > 300$~GeV. The successful description of data by the KS-gluon directly relies on including non-linear terms in the evolution; with those terms being absent, the description breaks down. \\

To observed slow-down of the growth with energy is one of the core
predictions of gluon saturation. We therefore are convinced that the
current study provides substantial evidence for the presence of the
saturation effects. We wish to note that a related observation has
already been made in \cite{Armesto:2014sma} on the level of  dipole
models. The current study substantiates this observation through
employing dipole cross-sections which are directly subject to linear
and non-linear QCD evolution. \\

Nevertheless it must be noted that the current study is not without deficits: the description is based on LO wave function and LO BK evolution (although supplemented with collinear resummations). To establish the observation made in this letter it is therefore necessary  to search for different observables which probe the low $x$ gluon in a similar kinematic regime and to increase further the theoretical accuracy of the underlying framework. Steps to address the  latter are  currently undertaken in \cite{Balitsky:2013fea,Hentschinski:2017ayz} (evolution equations) and \cite{Hentschinski:2018rrf,Hanninen:2017ddy,Balitsky:2010ze,Beuf:2017bpd} (determination of higher order corrections).

\section{Acknowledgments}
This work is partly supported by the Polish National Science Centre grant no. DEC-2017/27/B/ST2/01985 and
by COST Action CA16201 ”Unraveling new physics at the LHC through the precision frontier”.


\begin{thebibliography}{99}

\bibitem{Gelis:2015gza}
  F.~Gelis,
  Int.\ J.\ Mod.\ Phys.\ E {\bf 24} (2015) no.10,  1530008
  doi:10.1142/S0218301315300088
  [arXiv:1508.07974 [hep-ph]].

\bibitem{Gelis:2010nm}
  F.~Gelis, E.~Iancu, J.~Jalilian-Marian and R.~Venugopalan,
  Ann.\ Rev.\ Nucl.\ Part.\ Sci.\  {\bf 60} (2010) 463
  doi:10.1146/annurev.nucl.010909.083629
  [arXiv:1002.0333 [hep-ph]].


\bibitem{Gribov:1984tu}
  L.~V.~Gribov, E.~M.~Levin and M.~G.~Ryskin,
  Phys.\ Rept.\  {\bf 100} (1983) 1.
  doi:10.1016/0370-1573(83)90022-4



\bibitem{Balitsky:1995ub}
  I.~Balitsky,
  Nucl.\ Phys.\ B {\bf 463} (1996) 99
  doi:10.1016/0550-3213(95)00638-9
  [hep-ph/9509348].

\bibitem{JalilianMarian:1997jx}
  J.~Jalilian-Marian, A.~Kovner, A.~Leonidov and H.~Weigert,
  Nucl.\ Phys.\ B {\bf 504} (1997) 415
  doi:10.1016/S0550-3213(97)00440-9
  [hep-ph/9701284].


\bibitem{JalilianMarian:1997gr}
  J.~Jalilian-Marian, A.~Kovner, A.~Leonidov and H.~Weigert,
  Phys.\ Rev.\ D {\bf 59} (1998) 014014
  doi:10.1103/PhysRevD.59.014014
  [hep-ph/9706377].

\bibitem{Kovchegov:1999yj}
  Y.~V.~Kovchegov,
  Phys.\ Rev.\ D {\bf 60} (1999) 034008
  doi:10.1103/PhysRevD.60.034008
  [hep-ph/9901281].

\bibitem{GolecBiernat:1998js}
  K.~J.~Golec-Biernat and M.~Wusthoff,
  Phys.\ Rev.\ D {\bf 59} (1998) 014017
  doi:10.1103/PhysRevD.59.014017
  [hep-ph/9807513].
\bibitem{Kovchegov:2012mbw}
  Y.~V.~Kovchegov and E.~Levin,
  Camb.\ Monogr.\ Part.\ Phys.\ Nucl.\ Phys.\ Cosmol.\  {\bf 33} (2012) 1.
  doi:10.1017/CBO9781139022187



\bibitem{Albacete:2010pg}
  J.~L.~Albacete and C.~Marquet,
  Phys.\ Rev.\ Lett.\  {\bf 105} (2010) 162301
  doi:10.1103/PhysRevLett.105.162301
  [arXiv:1005.4065 [hep-ph]].

\bibitem{Lappi:2012nh}
  T.~Lappi and H.~Mantysaari,
  Nucl.\ Phys.\ A {\bf 908} (2013) 51
  doi:10.1016/j.nuclphysa.2013.03.017
  [arXiv:1209.2853 [hep-ph]].

\bibitem{Albacete:2018ruq}
  J.~L.~Albacete, G.~Giacalone, C.~Marquet and M.~Matas,
  Phys.\ Rev.\ D {\bf 99} (2019) no.1,  014002
  doi:10.1103/PhysRevD.99.014002
  [arXiv:1805.05711 [hep-ph]].

\bibitem{Albacete:2014fwa}
  J.~L.~Albacete and C.~Marquet,
  Prog.\ Part.\ Nucl.\ Phys.\  {\bf 76} (2014) 1
  doi:10.1016/j.ppnp.2014.01.004
  [arXiv:1401.4866 [hep-ph]].

\bibitem{Stasto:2018rci}
  A.~Stasto, S.~Y.~Wei, B.~W.~Xiao and F.~Yuan,
  Phys.\ Lett.\ B {\bf 784} (2018) 301
  doi:10.1016/j.physletb.2018.08.011
  [arXiv:1805.05712 [hep-ph]].

\bibitem{vanHameren:2019ysa}
  A.~van Hameren, P.~Kotko, K.~Kutak and S.~Sapeta,
  arXiv:1903.01361 [hep-ph].




\bibitem{Armesto:2014sma}
  N.~Armesto and A.~H.~Rezaeian,
  Phys.\ Rev.\ D {\bf 90}, no. 5, 054003 (2014)
  [arXiv:1402.4831 [hep-ph]].


\bibitem{Goncalves:2014wna}
  V.~P.~Goncalves, B.~D.~Moreira and F.~S.~Navarra,
  Phys.\ Rev.\ C {\bf 90}, no. 1, 015203 (2014)
  [arXiv:1405.6977 [hep-ph]].
\bibitem{Goncalves:2014swa}
  V.~P.~Goncalves, B.~D.~Moreira and F.~S.~Navarra,
  Phys.\ Lett.\ B {\bf 742}, 172 (2015)
  [arXiv:1408.1344 [hep-ph]].




\bibitem{Kowalski:2006hc}
  H.~Kowalski, L.~Motyka and G.~Watt,
  Phys.\ Rev.\ D {\bf 74}, 074016 (2006)
  [hep-ph/0606272].
\bibitem{Cox:2009ag}
  B.~E.~Cox, J.~R.~Forshaw and R.~Sandapen,
  JHEP {\bf 0906}, 034 (2009)
  [arXiv:0905.0102 [hep-ph]].



\bibitem{Ducloue:2016pqr}
  B.~Ducloué, T.~Lappi and H.~Mäntysaari,
  Phys.\ Rev.\ D {\bf 94} (2016) no.7,  074031
  doi:10.1103/PhysRevD.94.074031
  [arXiv:1605.05680 [hep-ph]].
\bibitem{Cepila:2018faq} 
  J.~Cepila, J.~G.~Contreras and M.~Matas,
  Phys.\ Rev.\ D {\bf 99}, no. 5, 051502 (2019)
  doi:10.1103/PhysRevD.99.051502
  [arXiv:1812.02548 [hep-ph]].



\bibitem{Jones:2013eda}
  S.~P.~Jones, A.~D.~Martin, M.~G.~Ryskin and T.~Teubner,
  J.\ Phys.\ G {\bf 41}, 055009 (2014)
  [arXiv:1312.6795 [hep-ph]].

\bibitem{Jones:2013pga}
  S.~P.~Jones, A.~D.~Martin, M.~G.~Ryskin and T.~Teubner,
  JHEP {\bf 1311}, 085 (2013)
  [arXiv:1307.7099].


\bibitem{Jones:2016icr} 
  S.~P.~Jones, A.~D.~Martin, M.~G.~Ryskin and T.~Teubner,
  J.\ Phys.\ G {\bf 44}, no. 3, 03LT01 (2017)
  doi:10.1088/1361-6471/aa56ea
  [arXiv:1611.03711 [hep-ph]].

\bibitem{Szczurek:2017uvc}
  A.~Szczurek, A.~Cisek and W.~Schafer,
  Acta Phys.\ Polon.\ B {\bf 48} (2017) 1207
  doi:10.5506/APhysPolB.48.1207
  [arXiv:1704.00444 [hep-ph]].


\bibitem{Bautista:2016xnp}
  I.~Bautista, A.~Fernandez Tellez and M.~Hentschinski,
  Phys.\ Rev.\ D {\bf 94} (2016) no.5,  054002
  [arXiv:1607.05203 [hep-ph]].

\bibitem{Hentschinski:2012kr}
  M.~Hentschinski, A.~Sabio Vera and C.~Salas,
  Phys.\ Rev.\ Lett.\  {\bf 110} (2013) no.4,  041601
  [arXiv:1209.1353 [hep-ph]];
  Phys.\ Rev.\ D {\bf 87} (2013) no.7,  076005
  [arXiv:1301.5283 [hep-ph]].

\bibitem{Celiberto:2018muu} 
  F.~G.~Celiberto, D.~Gordo Gómez and A.~Sabio Vera,
  Phys.\ Lett.\ B {\bf 786}, 201 (2018)
  doi:10.1016/j.physletb.2018.09.045
  [arXiv:1808.09511 [hep-ph]];
  A.~D.~Bolognino, F.~G.~Celiberto, D.~Y.~Ivanov and A.~Papa,
  Eur.\ Phys.\ J.\ C {\bf 78}, no. 12, 1023 (2018)
  doi:10.1140/epjc/s10052-018-6493-6
  [arXiv:1808.02395 [hep-ph]].





\bibitem{Kutak:2012rf}
  K.~Kutak and S.~Sapeta,
  Phys.\ Rev.\ D {\bf 86} (2012) 094043
  doi:10.1103/PhysRevD.86.094043
  [arXiv:1205.5035 [hep-ph]].

\bibitem{Kwiecinski:1997ee}
  J.~Kwiecinski, A.~D.~Martin and A.~M.~Stasto,
  Phys.\ Rev.\ D {\bf 56} (1997) 3991
  doi:10.1103/PhysRevD.56.3991
  [hep-ph/9703445].


\bibitem{Baranov:2007zza}
  S.~P.~Baranov,
  Phys.\ Rev.\ D {\bf 76}, 034021 (2007).


\bibitem{Brodsky:1980vj}
  S.~J.~Brodsky, T.~Huang and G.~P.~Lepage,
  ``The Hadronic Wave Function in Quantum Chromodynamics,''
  SLAC-PUB-2540.

\bibitem{Nemchik:1994fp}
  J.~Nemchik, N.~N.~Nikolaev and B.~G.~Zakharov,
  Phys.\ Lett.\ B {\bf 341}, 228 (1994)
  [hep-ph/9405355],
  J.~Nemchik, N.~N.~Nikolaev, E.~Predazzi and B.~G.~Zakharov,
  Z.\ Phys.\ C {\bf 75}, 71 (1997)
  [hep-ph/9605231],
  Z.\ Phys.\ C {\bf 75}, 71 (1997)
  [hep-ph/9605231].



\bibitem{Braun:2000wr}
  M.~Braun,
  Eur.\ Phys.\ J.\ C {\bf 16} (2000) 337
  doi:10.1007/s100520050026
  [hep-ph/0001268].

\bibitem{Chachamis:2015ona}
  G.~Chachamis, M.~De\`ak, M.~Hentschinski, G.~Rodrigo and A.~Sabio~Vera,
  JHEP {\bf 1509}, 123 (2015)
  [arXiv:1507.05778 [hep-ph]].




\bibitem{Chekanov:2002xi}
  S.~Chekanov {\it et al.} [ZEUS Collaboration],
  Eur.\ Phys.\ J.\ C {\bf 24}, 345 (2002)
  [hep-ex/0201043].
\bibitem{Chekanov:2004mw}
  S.~Chekanov {\it et al.} [ZEUS Collaboration],
  Nucl.\ Phys.\ B {\bf 695}, 3 (2004)
  [hep-ex/0404008].



\bibitem{Alexa:2013xxa}
  C.~Alexa {\it et al.} [H1 Collaboration],
  Eur.\ Phys.\ J.\ C {\bf 73}, no. 6, 2466 (2013)
  [arXiv:1304.5162 [hep-ex]].

\bibitem{Aktas:2005xu}
  A.~Aktas {\it et al.} [H1 Collaboration],
  Eur.\ Phys.\ J.\ C {\bf 46}, 585 (2006)
  [hep-ex/0510016].




\bibitem{TheALICE:2014dwa}
  B.~B.~Abelev {\it et al.} [ALICE Collaboration],
  Phys.\ Rev.\ Lett.\  {\bf 113}, no. 23, 232504 (2014)
  [arXiv:1406.7819 [nucl-ex]].
  S.~Acharya {\it et al.} [ALICE Collaboration],
  Eur.\ Phys.\ J.\ C {\bf 79}, no. 5, 402 (2019)
  doi:10.1140/epjc/s10052-019-6816-2
  [arXiv:1809.03235 [nucl-ex]].


\bibitem{Aaij:2013jxj}
  R.~Aaij {\it et al.} [LHCb Collaboration],
  J.\ Phys.\ G {\bf 40}, 045001 (2013)
  [arXiv:1301.7084 [hep-ex]];
  J.\ Phys.\ G {\bf 41}, 055002 (2014)
  [arXiv:1401.3288 [hep-ex]].
\bibitem{Aaij:2018arx} 
  R.~Aaij {\it et al.} [LHCb Collaboration],
  JHEP {\bf 1810}, 167 (2018)
  doi:10.1007/JHEP10(2018)167
  [arXiv:1806.04079 [hep-ex]].
\bibitem{Adloff:2000vm}
  C.~Adloff {\it et al.} [H1 Collaboration],
  Phys.\ Lett.\ B {\bf 483}, 23 (2000)
  [hep-ex/0003020].
\bibitem{Breitweg:1998ki}
  J.~Breitweg {\it et al.} [ZEUS Collaboration],
  Phys.\ Lett.\ B {\bf 437} (1998) 432
  [hep-ex/9807020].
\bibitem{Chekanov:2009zz}
  S.~Chekanov {\it et al.} [ZEUS Collaboration],
  Phys.\ Lett.\ B {\bf 680}, 4 (2009)
  [arXiv:0903.4205 [hep-ex]].
\bibitem{Aaij:2015kea}
  R.~Aaij {\it et al.} [LHCb Collaboration],
  JHEP {\bf 1509}, 084 (2015)
  [arXiv:1505.08139 [hep-ex]].

\bibitem{CMS:2016nct}
  CMS Collaboration [CMS Collaboration],
  ``Measurement of exclusive Y photoproduction in pPb collisions at $\sqrt{s_{_\mathrm{NN}}} = 5.02~\mathrm{TeV}$,''
  CMS-PAS-FSQ-13-009.
\bibitem{Sirunyan:2018sav} 
  A.~M.~Sirunyan {\it et al.} [CMS Collaboration],
  Eur.\ Phys.\ J.\ C {\bf 79}, no. 3, 277 (2019)
  doi:10.1140/epjc/s10052-019-6774-8
  [arXiv:1809.11080 [hep-ex]].



\bibitem{Weigert:2005us} 
  H.~Weigert,
  Prog.\ Part.\ Nucl.\ Phys.\  {\bf 55}, 461 (2005)
  doi:10.1016/j.ppnp.2005.01.029
  [hep-ph/0501087].

\bibitem{Bartels:1994jj} 
  J.~Bartels and M.~Wusthoff,
  Z.\ Phys.\ C {\bf 66}, 157 (1995).
  doi:10.1007/BF01496591


\bibitem{Ayala:2017rmh} 
  A.~Ayala, M.~Hentschinski, J.~Jalilian-Marian and
  M.~E.~Tejeda-Yeomans,  
  Phys.\ Lett.\ B {\bf 761}, 229 (2016)
  doi:10.1016/j.physletb.2016.08.035
  [arXiv:1604.08526 [hep-ph]];
  Nucl.\ Phys.\ B {\bf 920}, 232 (2017)
  doi:10.1016/j.nuclphysb.2017.03.028
  [arXiv:1701.07143 [hep-ph]];

\bibitem{Kotko:2017oxg} 
  P.~Kotko, K.~Kutak, S.~Sapeta, A.~M.~Stasto and M.~Strikman,
  Eur.\ Phys.\ J.\ C {\bf 77}, no. 5, 353 (2017)
  doi:10.1140/epjc/s10052-017-4906-6
  [arXiv:1702.03063 [hep-ph]].
\bibitem{Brodsky:1982gc}
  S.~J.~Brodsky, G.~P.~Lepage and P.~B.~Mackenzie,
  Phys.\ Rev.\ D {\bf 28} (1983) 228.




\bibitem{Rezaeian:2012ji} 
  A.~H.~Rezaeian, M.~Siddikov, M.~Van de Klundert and R.~Venugopalan,
  Phys.\ Rev.\ D {\bf 87}, no. 3, 034002 (2013)
  doi:10.1103/PhysRevD.87.034002
  [arXiv:1212.2974 [hep-ph]].

\bibitem{Bartels:2002cj} 
  J.~Bartels, K.~J.~Golec-Biernat and H.~Kowalski,
  Phys.\ Rev.\ D {\bf 66}, 014001 (2002)
  doi:10.1103/PhysRevD.66.014001
  [hep-ph/0203258];
H.~Kowalski and D.~Teaney,
  Phys.\ Rev.\ D {\bf 68}, 114005 (2003)
  doi:10.1103/PhysRevD.68.114005
  [hep-ph/0304189].





\bibitem{Balitsky:2013fea}
  I.~Balitsky and G.~A.~Chirilli,
  Phys.\ Rev.\ D {\bf 88} (2013) 111501
  doi:10.1103/PhysRevD.88.111501
  [arXiv:1309.7644 [hep-ph]].

\bibitem{Hentschinski:2017ayz}
  M.~Hentschinski, A.~Kusina, K.~Kutak and M.~Serino,
  Eur.\ Phys.\ J.\ C {\bf 78} (2018) no.3,  174
  doi:10.1140/epjc/s10052-018-5634-2
  [arXiv:1711.04587 [hep-ph]].
  M.~Hentschinski, A.~Kusina and K.~Kutak,
  Phys.\ Rev.\ D {\bf 94} (2016) no.11,  114013
  doi:10.1103/PhysRevD.94.114013
  [arXiv:1607.01507 [hep-ph]].
  O.~Gituliar, M.~Hentschinski and K.~Kutak,
  JHEP {\bf 1601} (2016) 181
  doi:10.1007/JHEP01(2016)181
  [arXiv:1511.08439 [hep-ph]].

\bibitem{Hentschinski:2018rrf}
  M.~Hentschinski,
  Phys.\ Rev.\ D {\bf 97} (2018) no.11,  114027
  doi:10.1103/PhysRevD.97.114027
  [arXiv:1802.06755 [hep-ph]].
  M.~Hentschinski and A.~Sabio~Vera,
  Phys.\ Rev.\ D {\bf 85} (2012) 056006
  [arXiv:1110.6741 [hep-ph]];
  M.~Hentschinski,
  Nucl.\ Phys.\ B {\bf 859} (2012) 129
  [arXiv:1112.4509 [hep-ph]];
  G.~Chachamis, M.~Hentschinski, J.~D.~Madrigal Martinez and A.~Sabio Vera,
  Nucl.\ Phys.\ B {\bf 876} (2013) 453
  [arXiv:1307.2591 [hep-ph]];
  Phys.\ Rev.\ D {\bf 87} (2013) no.7,  076009
  [arXiv:1212.4992];
  Nucl.\ Phys.\ B {\bf 861} (2012) 133
  [arXiv:1202.0649 [hep-ph]];
  M.~Hentschinski, J.~D.~Madrigal Mart\'inez, B.~Murdaca and A.~Sabio~Vera,
  Phys.\ Lett.\ B {\bf 735} (2014) 168
  [arXiv:1404.2937 [hep-ph]];
  Nucl.\ Phys.\ B {\bf 887} (2014) 309
  [arXiv:1406.5625 [hep-ph]];
  Nucl.\ Phys.\ B {\bf 889} (2014) 549
  [arXiv:1409.6704 [hep-ph]];

\bibitem{Hanninen:2017ddy}
  H.~Hänninen, T.~Lappi and R.~Paatelainen,
  Annals Phys.\  {\bf 393} (2018) 358
  doi:10.1016/j.aop.2018.04.015
  [arXiv:1711.08207 [hep-ph]].
\bibitem{Balitsky:2010ze}
  I.~Balitsky and G.~A.~Chirilli,
  Phys.\ Rev.\ D {\bf 83} (2011) 031502
  doi:10.1103/PhysRevD.83.031502
  [arXiv:1009.4729 [hep-ph]].
\bibitem{Beuf:2017bpd}
  G.~Beuf,
  Phys.\ Rev.\ D {\bf 96} (2017) no.7,  074033
  doi:10.1103/PhysRevD.96.074033
  [arXiv:1708.06557 [hep-ph]].
  R.~Boussarie, A.~V.~Grabovsky, D.~Y.~Ivanov, L.~Szymanowski and S.~Wallon,
  Phys.\ Rev.\ Lett.\  {\bf 119} (2017) no.7,  072002
  doi:10.1103/PhysRevLett.119.072002
  [arXiv:1612.08026 [hep-ph]].
  R.~Boussarie, A.~V.~Grabovsky, L.~Szymanowski and S.~Wallon,
  JHEP {\bf 1611} (2016) 149
  doi:10.1007/JHEP11(2016)149
  [arXiv:1606.00419 [hep-ph]].

\bibitem{Aaron:2009aa}
  F.~D.~Aaron {\it et al.} [H1 and ZEUS Collaborations],
  JHEP {\bf 1001} (2010) 109
  doi:10.1007/JHEP01(2010)109
  [arXiv:0911.0884 [hep-ex]].


\end{thebibliography}
\end{document}